\documentstyle[epsf,epsfig]{qcdparis}

\def\beq{\begin{equation}}
\def\eeq{\end{equation}}
\newcommand{\gesim}{\buildrel > \over {_\sim}}
\newcommand{\lesim}{\buildrel < \over {_\sim}}
\newcommand{\toa}{\buildrel A \over {_\to}}
\newcommand{\ie}{{\it i.e.}}
\newcommand{\eg}{{\it e.g.}}
\newcommand{\cf}{{\it cf.}}
\newcommand{\gev}{{\rm GeV}}
\newcommand{\order}[1]{${\cal O}(#1)$}
\newcommand{\morder}[1]{{\cal O}(#1)}
\newcommand{\eq}[1]{Eq. (\ref{#1})}
\newcommand {\ket}[1]{\vert{#1}\rangle}

\newcommand{\qpair}{$q\bar q$}
\newcommand{\jpsi}{$J/\psi$}

\begin{document}

\pagestyle{plain}

\title{Interactions  on  Nuclei$^{*,+}$}
\author{Paul Hoyer}
\affil{Nordita
\\ Copenhagen, Denmark}

 \abstract{I review hard photon initiated processes on nuclei. The
space-time development of the DIS reaction as viewed in the target rest
frame qualitatively describes the nuclear shadowing of quark and gluon
distributions, although it may be difficult to understand the
very weak $Q^2$ dependence of the low $x$ data. The current jet hadron
energy distribution at large $\nu$ is accurately independent of the target
size even at very small energy fractions $z \simeq 0.05$. Color transparency
is verified for vector meson $(J/\psi,\ \rho)$ production, but remains
enigmatic in quasiexclusive proton knockout processes. I emphasize the
importance of understanding short-range correlations in nuclei, as
manifested by subthreshold production and cumulative $x>1$ DIS processes.}

\resume{\ }

\twocolumn[\maketitle] \fnm{7}{Review talk given at
the Workshop on Deep Inelastic scattering and QCD, Paris, April 1995}
\fnm{6}{NORDITA - 95/65 P}

\section{Introduction}

The nucleus is a weakly bound (non-relativistic) state of protons and
neutrons. It would therefore appear that a hard scattering process
such as deep inelastic lepton scattering (DIS), with a coherence length $1/Q
\ll$ 1 fm, should give equivalent results for nuclear and free nucleon
targets. This view is false, as first demonstrated by the EMC
collaboration~\cite{emc} in 1982. Their data showed that the structure
function of an iron nucleus is not simply related to that of deuterium.

The EMC result led to a flurry of experimental and theoretical
activity, as documented in the comprehensive review of nuclear
effects in structure functions by Arneodo~\cite{arnrev}.  Today we know that
the nuclear structure function is not proportional to the nucleon one,
$F_2^A(x)\neq A F_2^N(x)$, for most values of the Bjorken scaling variable
$x= Q^2/2m_N \nu$, where $Q^2$ is the invariant momentum transfer squared
and $\nu$ is the energy of the photon in the target rest frame. A compilation
of newly reanalysed data from SLAC~\cite{slaceval} and NMC~\cite{nmceval} is
shown in \fref{nmcevalfig}, and a conventional nomenclature for the observed
nuclear effects is given in table 1.

\ffig{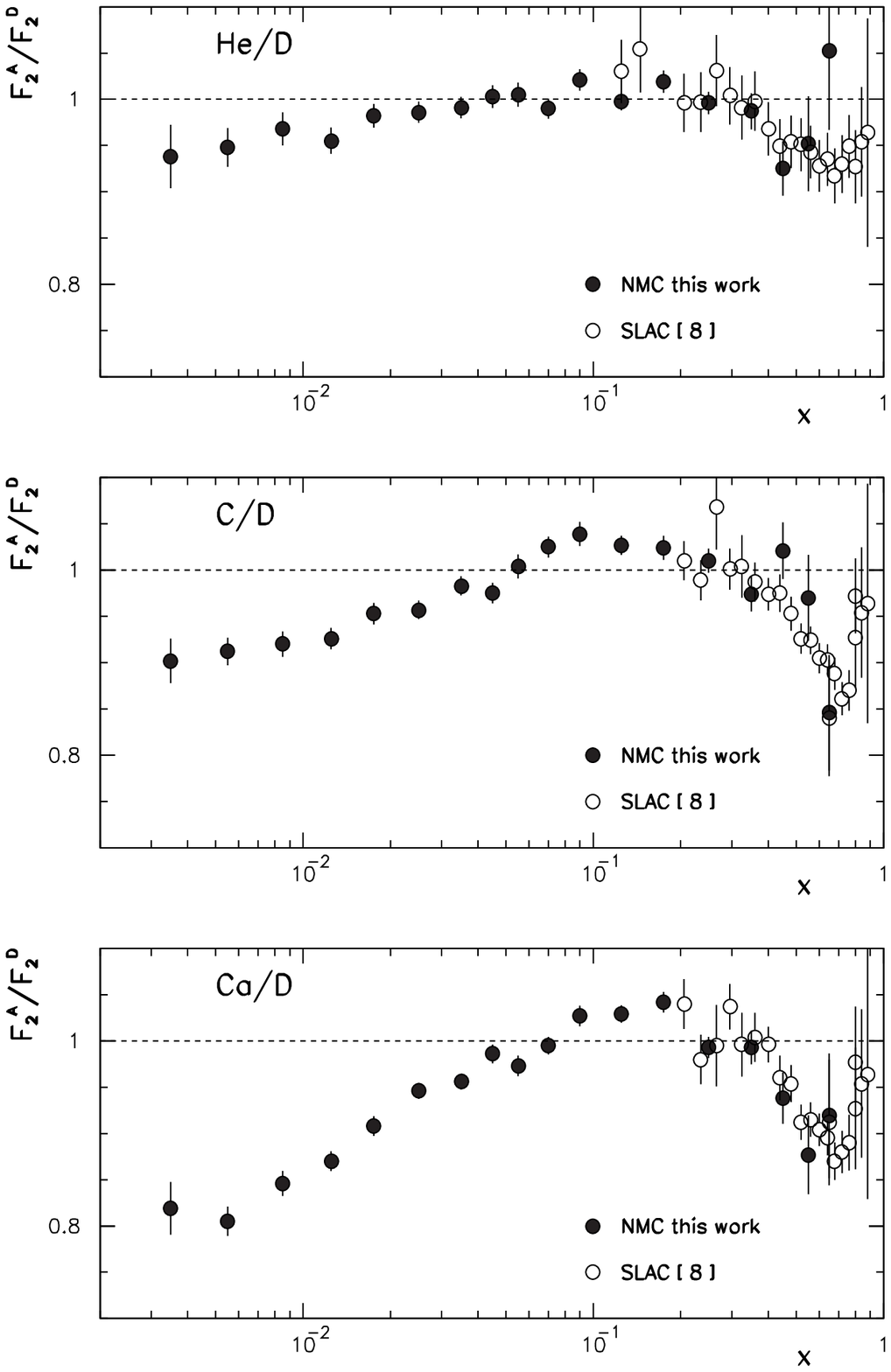}{120mm}{\em Compilation~\protect\cite{nmceval} of data
on the ratio of nuclear to deuterium structure functions for $He,\ C$ and $Ca$
targets.}{nmcevalfig}

\begin{table}[htb]
\begin{tabular}{|c|c|c|}\hline\rule[-3mm]{0mm}{8mm}
$F_2^A/AF_2^N$&$x-$range&Nuclear effect\\[1ex]\hline
$<1$&$x\leq 0.05$&Shadowing\\
$>1$&$.1\leq x \leq.2$&Anti-shadowing\\
$<1$&$.3\leq x \leq .8$&EMC effect\\
$>1$&$.8\leq x \leq 1$&Fermi motion\\
$\infty$&$1<x\leq A$&Cumulative effect\\\hline
\end{tabular}
\vskip5mm
\noindent{\bf Table 1.} Nomenclature of nuclear effects.
\end{table}

The underlying reason for the interesting properties of nuclei as measured
in DIS is that their nucleon constituents are relativistic bound states with
a rich internal structure of their own, which is resolved by the high $Q^2$
photons. The data implies that this nucleon structure is modified by the
nuclear environment. It should be stressed that the deviation of the ratio
$F_2^A/AF_2^N$ from unity is typically less than  20\ldots 30\% even for large
$A$~\cite{arnrev}. The gross features of the nuclear parton distributions are
thus similar to the nucleon ones, as expected.

Hence the quark distributions measured by DIS support the standard knowledge
that a nucleus may, to a first  approximation, be viewed as a collection of
weakly bound nucleons.

General arguments exist for the origin of the nuclear effects seen at small $x$
(shadowing, see below), and it is also clear that there will be important
effects at large values of $x$ (Fermi motion, cumulative effects). There is
still no consensus about the correct explanation of the EMC effect proper,
namely the suppression of the nuclear structure function for $.3 \leq x \leq
.8$.   Models have been proposed~\cite{arnrev} both from a hadronic (pions,
nuclear binding)  and a partonic (confinement radius, quark clusters) point of
view.   These two approaches are in principle complementary, but in practice
their relation to each other is unclear. Since the DIS measurement is highly
inclusive  (all partons except the observed quark are averaged over) the data
does not readily discriminate between the various proposals. A deeper
understanding requires comparisons of the model predictions also with less
inclusive measurements of the nuclear wave function.

It should be kept in mind that the nuclear effects observed in
DIS are for `average' nuclear configurations, which dominate in
the structure function. It is possible that the nuclear
effects are much larger if rare Fock states of the constituent nucleons
are selected. For example, the `cumulative' $x>1$ region is kinematically
accessible only in the case of nuclear targets.

In this review I discuss some recent topics involving hard
lepton scattering on nuclear targets, from the point of view of
a particle physicist. I shall argue the case that such reactions can
provide new insight on fundamental processes from two principal points of
view, as discussed in section 2.  In section
\ref{spacetime} I review the space-time picture of DIS in the target rest
frame, and discuss some developments since the review ~\cite{arnrev}. Nuclear
effects on quark jet hadronization are covered in section \ref{qhadr}, and in
section \ref{ctdis} I discuss color transparency. Section \ref{cumsec} is
devoted to rare, high density fluctuations in nuclei.

A quantitative treatment of nuclear effects generally requires detailed
nuclear modelling, the validity of which is difficult for a particle physicist
to assess. Here I shall mainly emphasize general, model-independent trends, and
refer to the original papers for the motivations of specific assumptions.

\section{The two uses of nuclear targets}

It is helpful to note that hard interactions in nuclei can be used
in two complementary ways, either to give information about the time
evolution of the produced states, or to investigate the properties of the
nucleus itself.

\subsection{The nucleus as a femtovertex detector}

The nucleus may (ideally) serve as the smallest conceivable
vertex detector. Following (or preceding) a hard collision on a quark or gluon
in the nucleus,  the rest of the nucleus serves as a medium for detecting
secondary  interactions of the produced partons.
Examples are the hadronization of the recoil quark in DIS (section
\ref{qhadr}), and the  propagation of the  $q\bar q$  state in $eA \to e\rho A$
(section \ref{vmct}).  The usefulness of the femtodetector depends on
how well  it can be calibrated, i.e., on our understanding  of the secondary
interactions.  I shall discuss below some general features - more
will be learnt in future experimental and theoretical studies.

\subsection{Study of rare nuclear configurations}

Hard interactions on nuclei may be used to study short range correlations
in the nuclear wave function. In this case the nucleus is the object
of study rather than the detector. For structures much smaller than 1 fm we are
in the domain of perturbative QCD and should be able to calculate the
probability of rare, shortlived nuclear configurations starting from the
longlived ones.   An example is provided by the `cumulative' $(x>1)$
region of DIS, where several nucleons deliver  their momentum to a
single quark or other compact partonic subsystem (section \ref{cumsec}).

\subsection{Combinations of the above two uses of the nucleus}

Sometimes we wish to make a combined use of the nucleus, selecting a rare
short-range nuclear configuration and then using the rest of the nucleus as a
detector (analogously to what was done in bubble chambers). An example of this
is quasielastic $ep$ scattering in a nucleus, observed through the nucleon
knock-out reaction $eA \to ep(A-1)$. This process is of considerable interest
for studying color transparency (section \ref{pct}). The struck constituent
proton is selected to be in a rare, compact configuration, whereas the
remainder of the nucleus serves to measure the rescattering cross section of
this `small proton'. An essential assumption needed for a color transparency
interpretation is that the probability to find the small-sized proton is
independent of the nuclear size $A$.

\section{The space-time picture of high energy scattering} \label{spacetime}

The general features of the time development of high energy $eA$ scattering
can be established using only Lorentz invariance and the
uncertainty principle. For understanding the specifically nuclear effects it
is best to view the scattering in the target rest frame, where we have an
intuitive understanding of nuclear structure.

\ffig{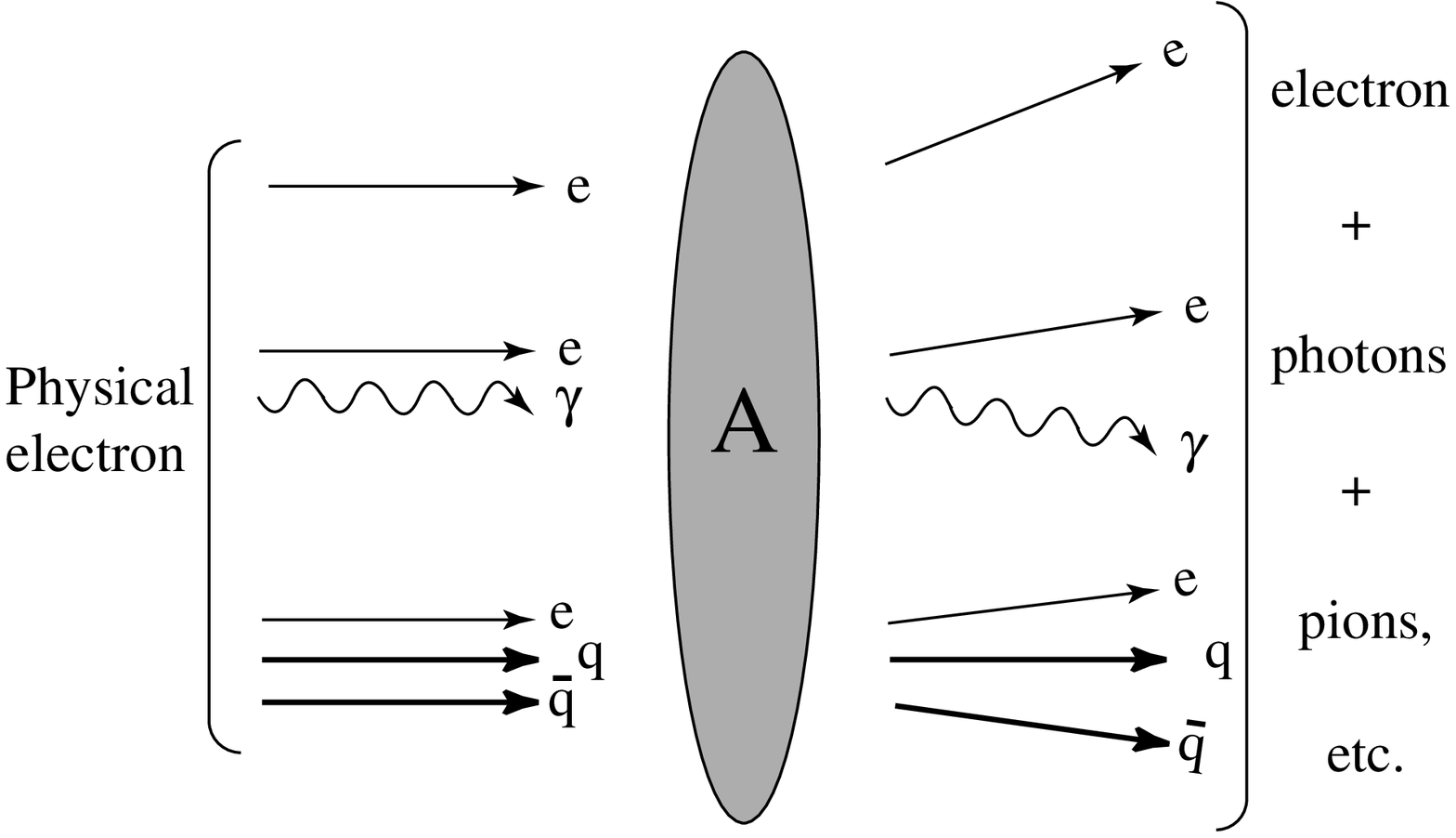}{50mm}{\em Space-time picture of $eA$ scattering. Three
Fock states, $\vert{e}\rangle,\ \vert{e\gamma}\rangle$ and $\vert{eq\bar
q}\rangle$ are shown to scatter elastically at high energy.} {eadiag}

The incoming physical electron state (see \fref{eadiag}) can, at a
given instant  of time, be expanded in terms of its (bare) Fock states
\beq
\ket{e}_{phys}=\psi_e\ket{e}+\psi_{e\gamma}\ket{e\gamma}+\psi_{eq\bar
q}\ket{eq\bar q} + \ldots \label{fock}
\eeq
The amplitudes $\psi_i$ depend on the kinematic variables
describing the states $\ket{i}$, and have a time dependence
$\exp(-i E_i t)$, where $E_i = \sum_i \sqrt{m_i^2+{\vec p_i^2}}$ is the free
(kinetic) energy of the partons. (Note that since the Fock expansion is at a
fixed time $t$, energy is not conserved and $E_i$ differs from $E_e$, the
energy of the physical electron.) The `lifetime' $\tau_i \simeq 1/(E_i-E_e)$ of
a Fock state
$\ket{i}$ is given by the time interval after  which the relative phase
$\exp[-i(E_i-E_e)]$ is significantly different from unity.  There is a
continuous  mixing of Fock states, with state $\ket{i}$ mixing at a rate
$1/\tau_i$.

At high electron energies $E_e$ the lifetimes $\tau_i \propto E_e$
are dilated by the Lorenz factor.  If $\ket{i}$ contains partons
of mass $m_j$, energy fraction $x_j$ and transverse momenta $p_{\perp j}$
we have explicitly
\beq
E_i-E_e \simeq \frac{1}{2E_e}\left(\sum_j \frac{m_j^2+p_{\perp j}^2}{x_j}-m_e^2
\right) \label{ediff}
\eeq
In typical soft collisions the Fock states in \eq{fock} form long before the
electron  arrives at the nucleus, and they live long after their passage.
There is no time within the nucleus to form a new Fock state
(\eg, by radiating a gluon), unless its lifetime is of the order of the nuclear
radius, $\tau \lesim R_A$. Hence the scattering inside the nucleus
typically is {\em diagonal} in the Fock basis: $\ket{e} \toa \ket{e}$,
$\ket{e\gamma} \toa \ket{e\gamma}$, and so on, as indicated in \fref{eadiag}.

The transverse velocities $v_{\perp j}=p_{\perp j}/x_jE_e$ are typically
small at large $E_e$.  Hence the impact parameters
(transverse coordinates) of all partons are preserved.

The only effect of the nucleus on the Fock states
is then to impart transverse momenta via elastic scattering.  This is enough to
upset the delicate balance in the electron Fock state mixing so that the
asymptotic state which emerges (long after nuclear traversal) can
contain a shower of photons and hadrons.

\subsection{DIS as seen in the target rest frame} \label{disrest}

In hard collisions such as deep inelastic scattering, where the momentum
transfer is commensurate with the incoming energy, the reaction times are short
and the above picture needs to be refined. The usual interpretation of DIS as a
measurement of the target structure functions is simple in the frame where
also the target has high momentum (or, equivalently, in terms of light cone
coordinates). The nuclear target effects, on the other hand, are easier to
discuss in the nuclear rest frame~\cite{ioffer,stref,niza,dbp}.

Deep inelastic scattering  $eA \to e + X$ is characterized by a
large electron energy loss $\nu$ (in the target rest frame) and an invariant
momentum transfer  $q^2 = - Q^2$ between the incoming and outgoing electron
such that $x=Q^2/2m_N\nu$ is fixed.  In terms of Fock states, the electron
first emits a photon ($\ket{e} \to \ket{e\gamma}$)  with $E_\gamma=\nu$ and
$p_{\perp\gamma}^2=Q^2(1-\nu/E_e)$. The energy difference
\beq
E_{e\gamma}-E_e =\frac{Q^2}{2\nu} = m_Nx \label{esplit}
\eeq
is {\em fixed} in the Bjorken limit, implying that the $\ket{e\gamma}$ state
typically travels a distance
\beq
2L_I = \frac{2\nu}{Q^2} = \frac{1}{m_Nx}= \left\{ \begin{array}
{rl}1\ \rm fm&\mbox{\rm for $x=.2$}\\
200\ \rm fm&\mbox{\rm for $x=.001$}\\
\end{array} \right. \label{ioffe}
\eeq
defined as two `Ioffe lengths'~\cite{ioffer}  $L_I$.  The factor two is
conventional, and motivated by the fact that the photon
still splits into a  $q\bar q$  pair before interacting in the nucleus:
$\ket{e\gamma} \to \ket{eq\bar q}$. If the antiquark $\bar q$ carries a
fraction
$\alpha$ of the photon energy, this transition involves another energy
difference
\beq
E_{\gamma}-E_{q\bar q} = \frac{1}{2\nu}\frac{m_q^2+p_{\perp
q}^2}{\alpha(1-\alpha)} = {\cal O}(Q^2/2\nu) \label{fsplit}
\eeq
which should be as big as the previous one, \eq{esplit}, to have time to happen
during the life-time of the $e\gamma$ state.  There are two principal
ways in which the large energy difference indicated in \eq{fsplit} can arise.

\subsubsection{Parton model regime} \label{parsec}

\beq
\alpha = {\cal O}(1/Q^2),\ \ \ \ p_{\perp q}= {\cal O}(\Lambda_{QCD})
\label{parton}
\eeq

The energy of the $\bar q$ is finite in the target rest
frame\footnote{Equivalently, we may have $1-\alpha = {\cal O}(1/Q^2)$ and a
finite quark energy.}: $\alpha\nu={\cal O}(1/x)$. Its transverse velocity
$v_{\perp}(\bar q)= p_{\perp q}/\alpha\nu = {\cal O}(x)$,
hence during the life-time of the $q\bar q$ state it expands a transverse
distance
\beq
r_\perp(q\bar q) = v_\perp L_I = \morder{1\ {\rm fm}}  \label{qsize}
\eeq
provided $m_q \lesim \Lambda_{QCD}$.

Depending on the value of $x$,  the asymmetric  \qpair\  pair is created
either {\em (a)} in the nucleus $(x\gesim 0.1,\ L_I\lesim 1\ {\rm fm})$ or
{\em (b)} well before the nucleus $(x\lesim 0.01,\ L_I\gesim 10\ {\rm fm})$.
The antiquark interacts in the nucleus with  a large cross-section, as
dictated  by its large transverse spread in \eq{qsize}.  In case (a)
$\sigma_{DIS}(eA)\propto A$ while in case (b) the $\bar q$ scatters on
the nuclear surface and  $\sigma_{DIS}(eA)\propto A^{2/3}$.
In either case the fast bare quark begins to radiate soft gluons
and hadronize only well after the nucleus.  In higher orders of
$\alpha_s(Q^2)$  the quark may radiate hard gluons before or inside the
nucleus,  but this radiation is independent of the nucleus and hence does not
change the A-dependence of $\sigma_{DIS}(eA)$.
To leading order in $1/Q^2$ the fast quark only experiences soft elastic
scattering in the nucleus (\cf\ \fref{eadiag}).

\subsubsection{Gluon scattering regime} \label{glusec}

The other possibility of achieving a large energy difference in \eq{fsplit} is
\beq
\alpha = {\cal O}(1/2),\ \ \ \ p_{\perp q}= {\cal O}(Q) \label{gluon}
\eeq
Now the quark and antiquark share the photon energy roughly equally.  The
transverse velocity is
$v_{\perp}= \morder{Q/\nu} = \morder{x/Q}$,
implying a small transverse size of the pair, $r_\perp(q\bar q) =
\morder{1/Q}$. The quark pair has a small interaction cross section in the
nucleus,  but may (within the lifetime of the pair, and at the price of a
coupling constant $\alpha_s(Q^2)$) interact by emitting a  gluon of energy
fraction \order{1/Q^2}. At small $x$, the gluon is created before arrival at
the nucleus and scatters off the nuclear surface, which again results in
shadowing. If one assumes that the scattering cross section for a gluon on
the nucleus is larger than that of an (anti)quark, this would imply that the
shadowing effect is larger in this regime than in the parton model case
(\ref{parton}). So far there is little direct experimental information
available on the shadowing of the gluon structure function.

It is important to notice that the wee (anti)quark in case (\ref{parton}) and
the wee gluon in case (\ref{gluon}) have finite momenta of \order{1/x} in the
target rest frame. This makes it possible that they are interpreted as
belonging to the target wave function in the more familiar frame where the
target has large momentum. The parton subprocesses corresponding to cases
(\ref{parton}) and (\ref{gluon}) are then $\gamma^*q \to q$ and $\gamma^*g \to
q\bar q$, respectively.

\subsection{Shadowing and $\sigma_L/\sigma_T$}

A well-known prediction of the parton model is that DIS
is dominated by the scattering of transverse photons, \ie,
$R=\sigma_L/\sigma_T=0$ (the Callan-Gross relation). In the target rest frame
picture (\ref{parton}) this is a consequence of the fact that only transverse
photons readily split asymmetrically into $q\bar q$ pairs where one of the
quarks carries wee momentum (see section \ref{vmct}). For the gluon scattering
case (\ref{gluon}) there is no similar restriction, hence
$R=\morder{\alpha_s}$. For scattering on nuclei, it is possible that
shadowing affects $\sigma_L$ differently from $\sigma_T$ (\eg, if the gluon
structure function is more shadowed than the quark one~\cite{niza,dbp} or
due to higher twist effects induced by Fermi motion~\cite{mil}). The (scant)
available data on the nuclear dependence of R in the shadowing region
suggests at most a small effect.  An earlier NMC result $R^{Ca}-R^C= 0.027\pm
0.026\pm 0.020$ (for $.007 < x < .2$) was consistent with no effect.
Preliminary data presented by  the NMC at this meeting~\cite{muc} shows a
small positive result,   $R^{Sn}-R^C= 0.031\pm 0.017$, consistent with a
larger shadowing for gluons.

\subsection{$Q^2$ Dependence of Shadowing}

In recent years, extensive data on nuclear structure functions in the
shadowing region has been obtained in particular by the
NMC~\cite{nmceval,nmcshad,nmcrat} and E665~\cite{e665shad,e665lowx}
collaborations. The distributions extend to very low $x$ as seen in
\fref{lowxshadow} from Ref.~\cite{e665lowx}.

\ffig{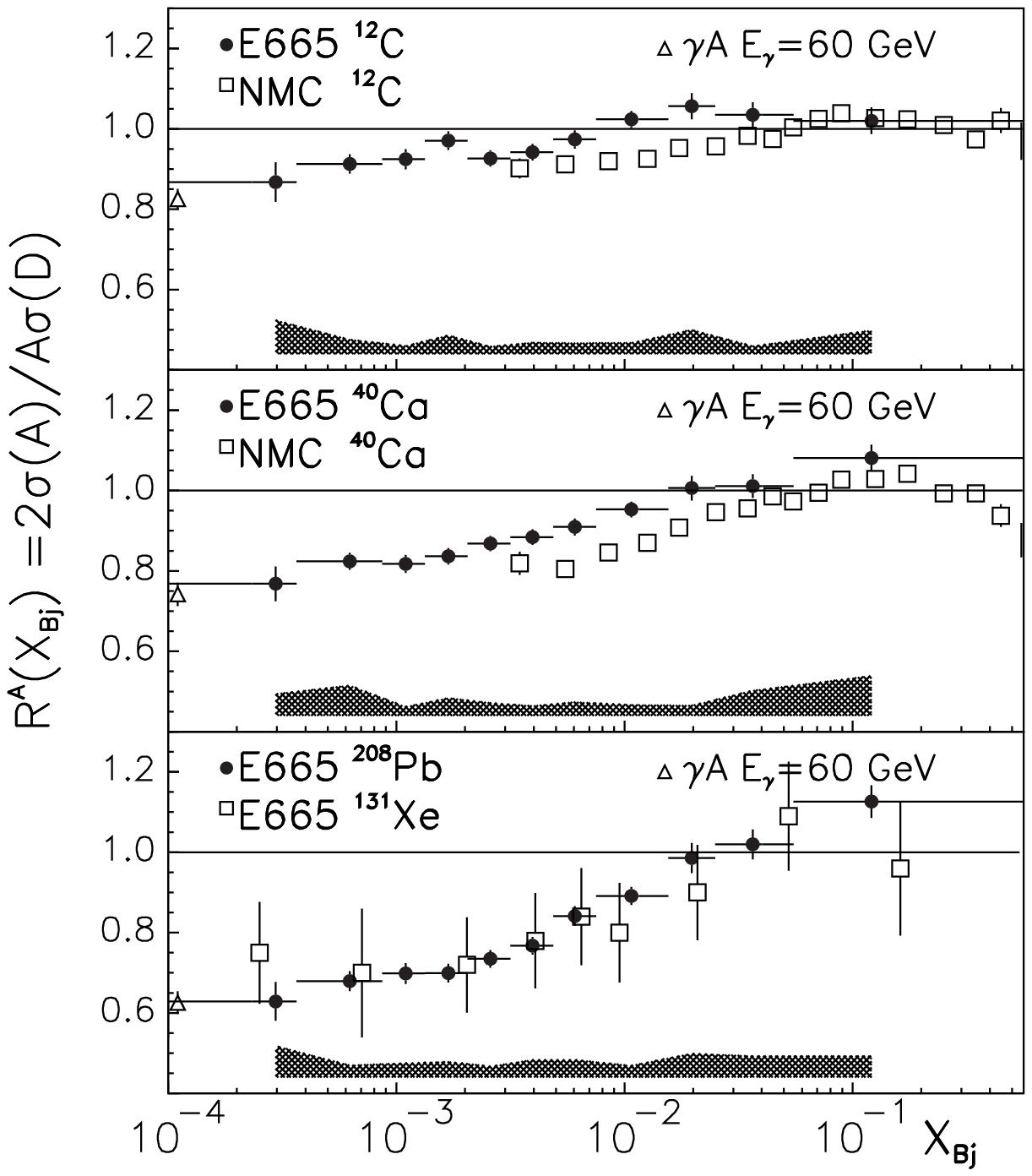}{100mm}{\em E665~\protect\cite{e665lowx} and
NMC~\protect\cite{nmcshad} data on shadowing in the structure function at low
$x$.}{lowxshadow}

When the nuclear DIS cross section is parameterized as
$\sigma(\gamma^*A)=A^{\alpha(x)} \sigma(\gamma^*N)$, E665
finds~\cite{e665lowx} that the exponent $\alpha(x)$ decreases with $x$ from
$\alpha(0.05)\simeq 0$ to $\alpha(0.002) = 0.906 \pm 0.006$ and remains
consistent with the latter value for $0.0003 \le x \le 0.002$.

The kinematics of the fixed target data is such that
$Q^2$  decreases with $x$, with $\langle Q^2 \rangle \lesim 0.5$ GeV$^2$ for $x
\le 0.002$ in E665. Hence the data should join smoothly with the real
photoproduction $(Q^2 = 0)$  data for $x \to 0$, as is in fact observed in
\fref{lowxshadow}.

Due to the low values of $Q^2$ one might expect the shadowing
effect  to show some residual $Q^2$ dependence.  The published
NMC~\cite{nmcshad,nmcrat} and  E665~\cite{e665lowx} data nevertheless shows
that the shadowing is at most weakly dependent on $Q^2$ at fixed $x$, see
\fref{qdep}.

\ffig{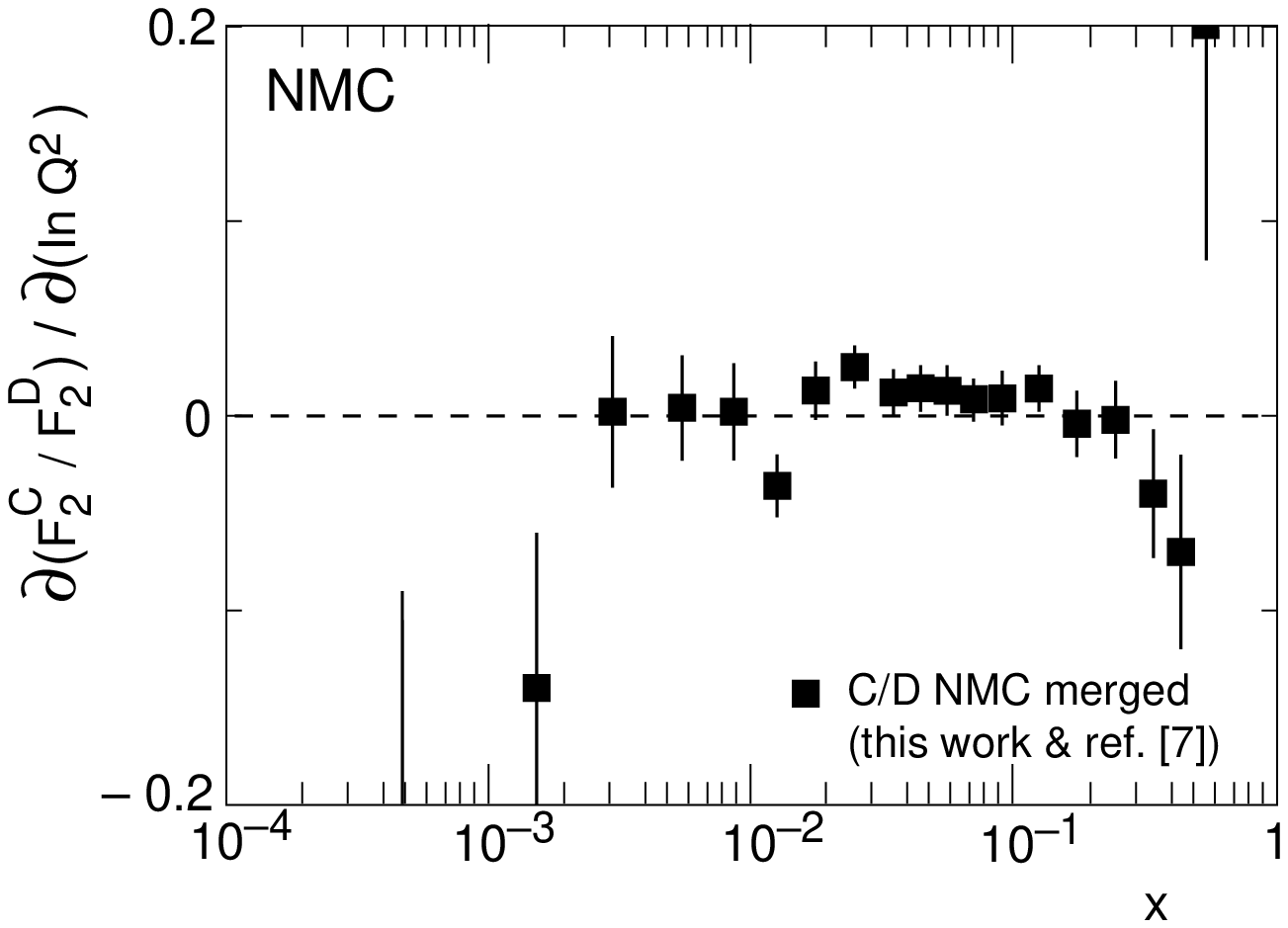}{60mm}{\em The dependence of $F_2^C/F_2^D$ on $\log Q^2$
as measured by NMC~\protect\cite{nmcrat}.}{qdep}

\subsection{Theoretical models of shadowing in DIS}

Quantitative models for describing the shadowing effect, based on the
target rest frame view outlined in section \ref{disrest}, have been
constructed by several authors~\cite{stref,niza,blu,bakw,meth,piwe}. In
order to properly describe also the very small $x$, low $Q^2$ data the
contribution of vector meson production has been taken into account. With
plausible parametrizations of the \qpair\ cross section on nuclei the
shadowing effect can be adequately described. The models~\cite{meth,piwe}
give a somewhat larger $Q^2$ dependence for the shadowing effect than is
observed in the present data. A prediction for the $Q^2$ dependence of the
$F_2^{Sn}/F_2^C$ ratio from Ref.~\cite{meth} is shown in \fref{sncdep}. The
model of Ref.~\cite{piwe} gives a very similar prediction. Preliminary NMC
data reported at this meeting~\cite{muc} does indicate a positive $Q^2$
dependence for this ratio.

\ffig{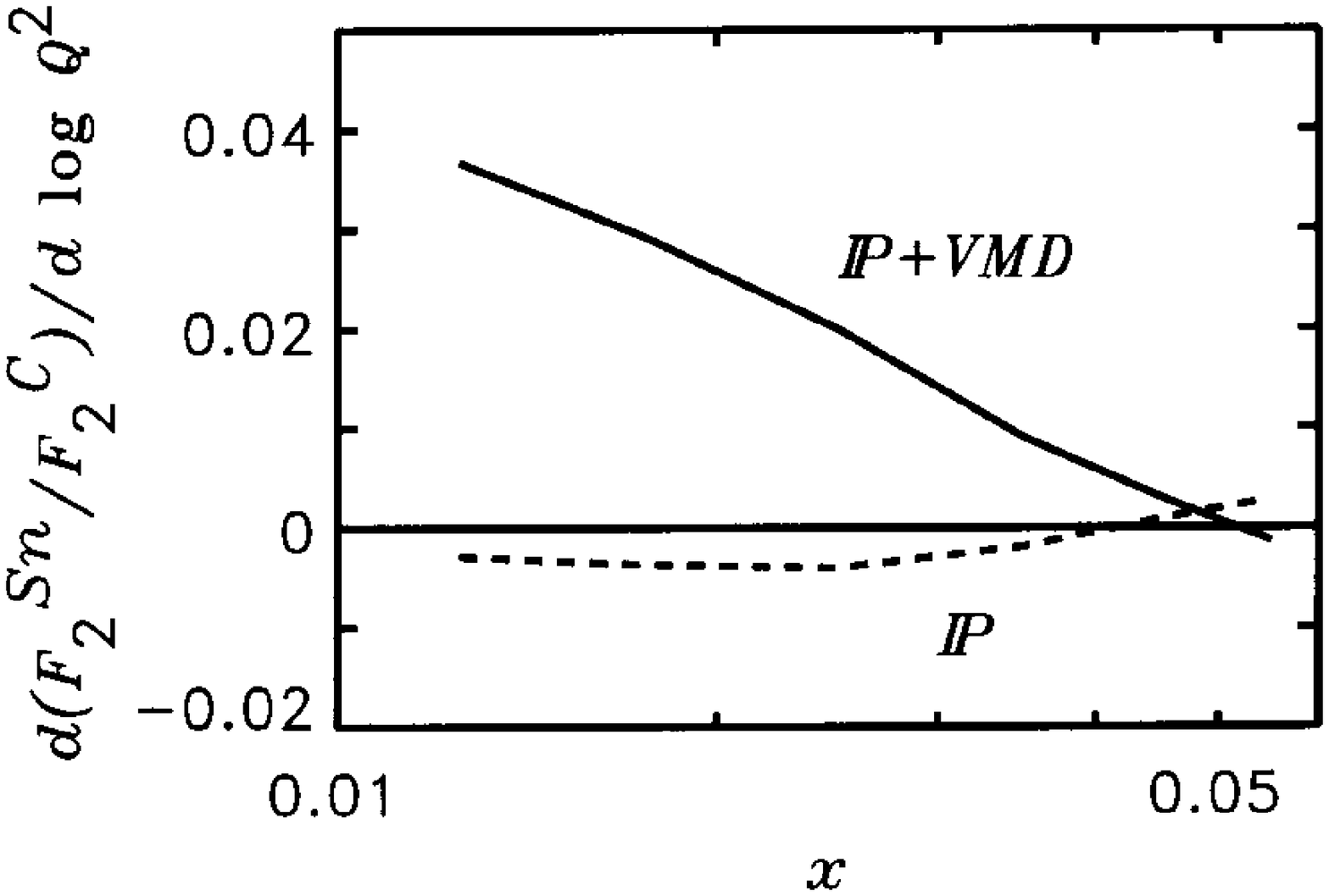}{60mm}{\em The dependence of $F_2^{Sn}/F_2^C$ on $\log
Q^2$ according to~\protect\cite{meth}. The solid curve is the full result,
while the dashed one shows the Pomeron contribution only.}{sncdep}

At very low values of $x$ the parton densities of heavy nuclei can become
large enough for parton recombination to occur in the $Q^2$
evolution~\cite{muqi}. This gives rise to another type of shadowing effect
which is not, however, expected to be relevant in the $x$ and $Q^2$
range of the present data.

An explanation of the 'anti-shadowing' nuclear enhancement
effect observed for $0.1 \lesim x \lesim 0.2$ has been offered by
Brodsky and Lu~\cite{blu}, who use a Regge model for $\bar q A$
scattering  (recall that the $\bar q$ momentum is of \order{1/x} and hence
large at low $x$). They find (with a suitable choice of the Regge parameters)
that the enhancement can be  understood as due to interference between the
leading (Pomeron)  and secondary (meson)  Regge exchanges.

It was recently proposed~\cite{kp} that the shadowing effect should
scale as a function of the number of gluons $n(x,Q^2,A)$ which interact with
the $q\bar q$ fluctuation during its lifetime. This number was estimated as
\beq
n(x,Q^2,A)=\frac{\langle\sigma^2(\rho)\rangle}{4\langle\sigma(\rho)\rangle}
\langle T(b)\rangle F_A^2(q) x^{-\Delta_P(Q^2)} \label{ngl}
\eeq
Here $\rho$ is the transverse size of the $q\bar q$ pair, $\sigma(\rho,x)=
\sigma(\rho)x^{-\Delta_P(Q^2)}$ is the interaction cross section of the quark
pair with a nucleon, $T(b)$ is the nuclear density profile in impact parameter
space and $F_A(q)$ is the nuclear longitudinal form factor. Assuming specific,
physically motivated forms for the quantities appearing in \eq{ngl}, the
scaling prediction seems to be in good agreement with the available data, as
shown in
\fref{kopfig}.

\ffig{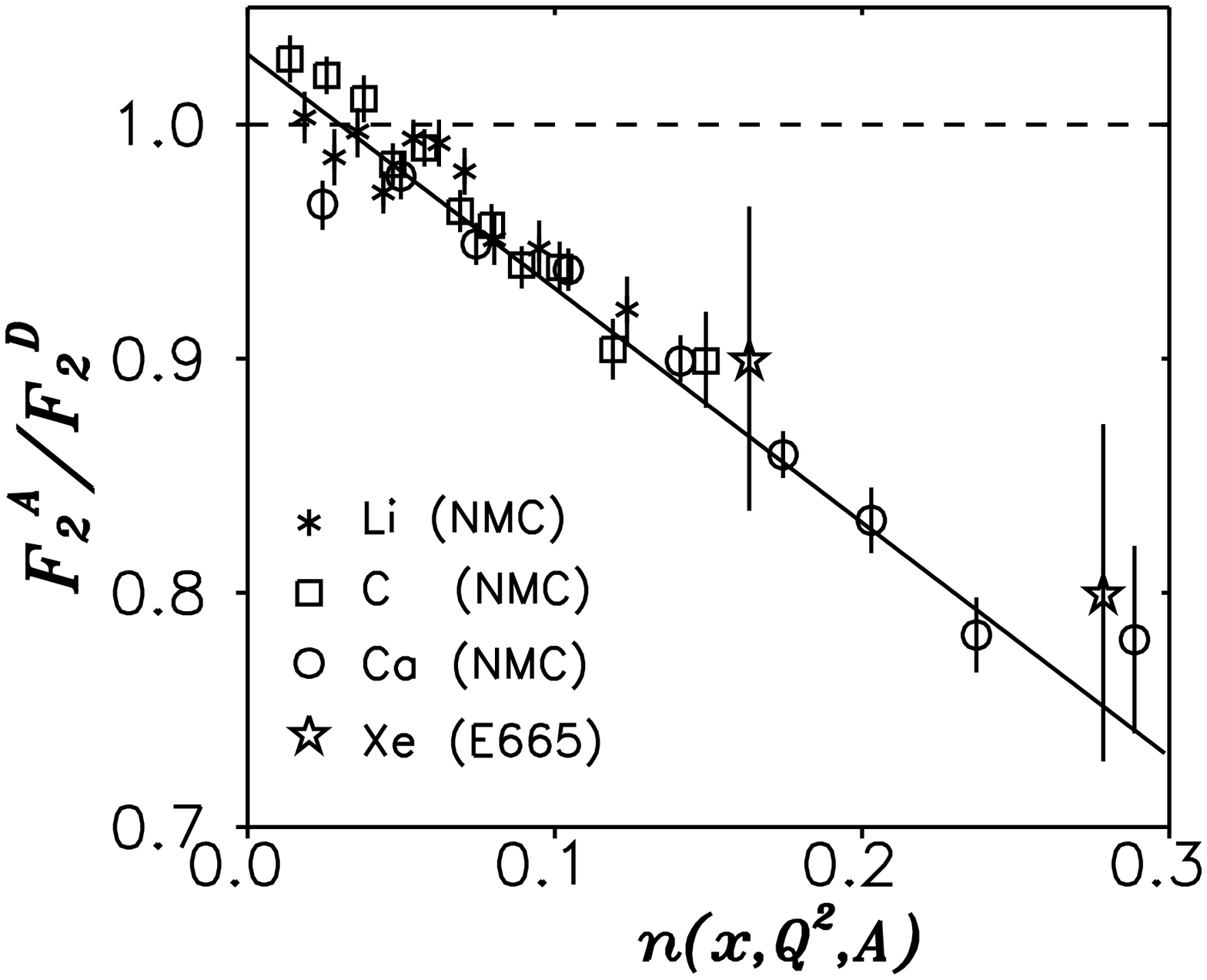}{70mm}{\em A scaling prediction of shadowing
compared with data~\protect\cite{kp}. The scaling variable $n(x,Q^2,A)$ is
defined by \protect\eq{ngl}.}{kopfig}

\subsection{Structure functions in configuration space}

Due to the highly inclusive nature of DIS, the measurements provide
only limited possibilities of testing detailed model assumptions. In view
of this it is interesting to note that the structure functions can, in a
model independent way, be studied also in coordinate space~\cite{bgm}. The
coordinate and momentum space descriptions of the structure functions contain
equivalent information and are related by a Fourier transform.  For example,
for the valence quark structure function the relation is
\beq
q_{val}(z,Q^2) = \int_0^1 dx\,\cos(xz)q_{val}(x,Q^2) \label{qfour}
\eeq
where $x$ is the standard Bjorken energy fraction and
$z = m_N L_I$ ($L_I$ = Ioffe length, see \eq{ioffe}) measures the longitudinal
light cone size of the quark distribution.  The  $Q^2$ evolution of
$q_{val}(z,Q^2)$ is given implicitly by \eq{qfour} in terms of that for
$q_{val}(x,Q^2)$. The $Q^2$ evolution equation can also be written
directly in coordinate space.

It may be worthwhile to study the nuclear effects on the structure
functions also in coordinate space.  This will give a model independent
characterization of how the spatial distribution of quarks in nucleons
is modified in the nuclear environment.

\section{Hadronization of the quark jet} \label{qhadr}

In our discussion of the space-time picture of DIS, we noted that the
photon splits asymmetrically into a $q\bar q$ pair, such that one of the quarks
takes almost all the photon momentum. This quark (which is the final state
struck quark in the infinite momentum frame picture) forms the `current jet' of
hadrons. For nuclear targets, the fast quark must first penetrate the nucleus.
At sufficiently high hadron energy $E_h \gesim \morder{R_A \langle
p_\perp^2\rangle}$, where
$\langle p_\perp^2\rangle$ is a hadronization energy scale
of \order{\Lambda_{QCD}^2}, the hadronization will (due to time dilation) start
only after the quark has penetrated the nucleus. The hadronization should then
be {\em independent} of  the target size $A$.

Note that there can be hard gluon emission at a short time scale inside the
nucleus, when the quark is produced at high virtuality. Such emission is
associated with the hard vertex and independent of the the nuclear size.
Furthermore, the quark is intially `bare' (unaccompanied by soft gluon
radiation) and at large $\nu$ has no time to form a soft gluon cloud while
inside the nucleus. Hence there is no
$A$-dependent energy loss~\cite{bh}, only elastic quark scattering as
shown in \fref{eadiag}.

\ffig{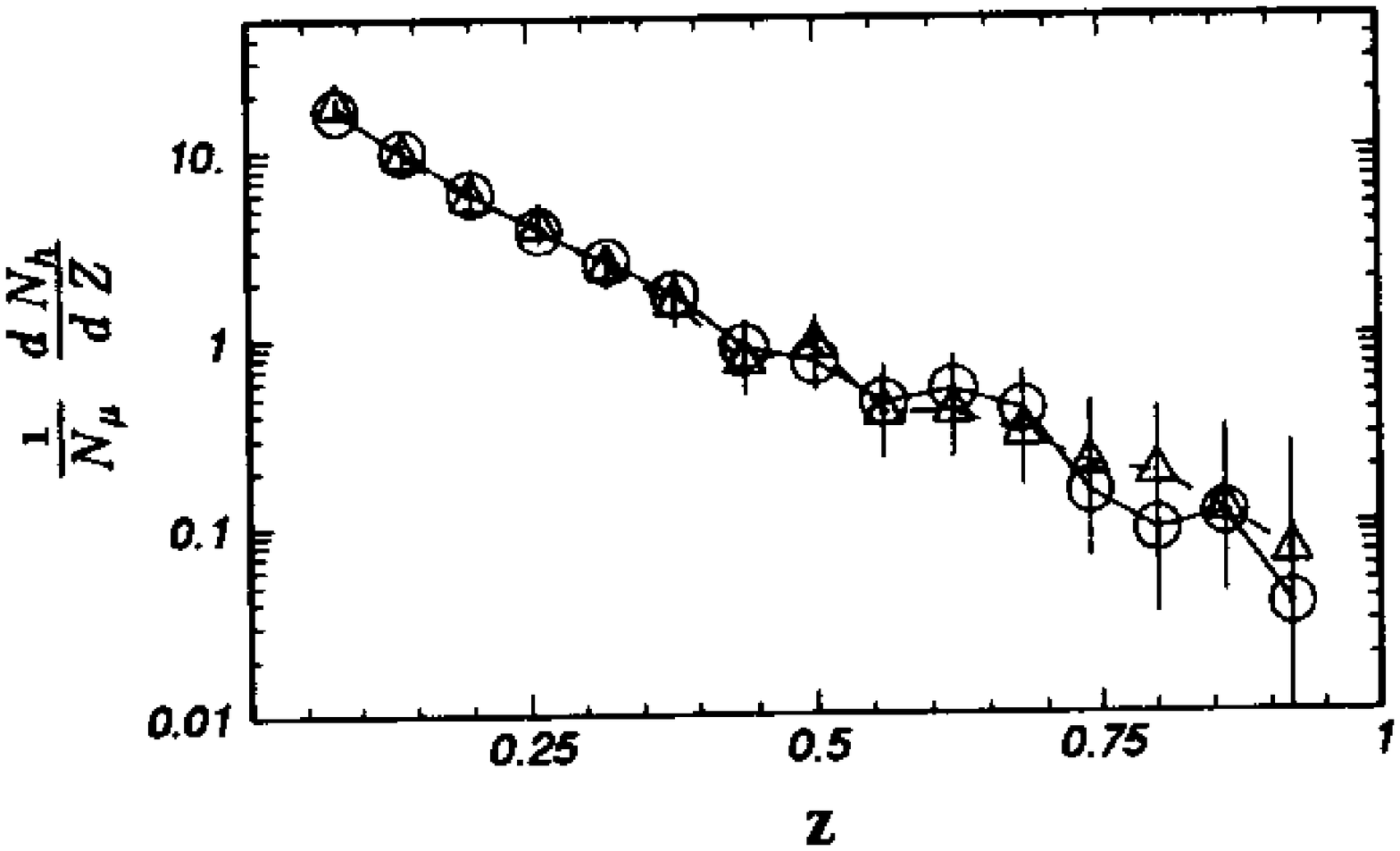}{55mm}{\em Hadron distributions in
$\mu A$ scattering~\protect\cite{e665zdist} as a function of the hadron
fractional laboratory energy $z=E_h/\nu$, for $A=D_2$ (circles) and $A=Xe$
(triangles). The events satisfy the constraints $x<0.005$ and
$Q^2<1\ \gev^2$, a kinematic region where a strong shadowing effect
is observed in the DIS cross section.}{zscaling}

The data agrees well with this simple picture. As an example, I show in
\fref{zscaling} the 490 GeV $\mu A$ data of E665~\cite{e665zdist} on the
inclusive hadron momentum spectrum in the current jet. The hadron distribution
is plotted as a function of the fractional energy $z$ that the hadron carries
of the virtual photon energy $\nu$. There is no observable difference between
the distribution measured for a heavy target $(Xe)$ compared to that for a
light target $(D_2)$, in the whole measured range $.05 \le z \le .95$. If the
fast quark suffered energy loss in the nucleus one would, on the contrary,
expect that the $z$-distribution for $Xe$ would be steeper than that for
$D_2$. The data was interpreted~\cite{e665zdist} as an upper limit of 1.7 mb
(90 \% c.l.) for the effective nuclear rescattering cross section of the fast
quark.

The data in \fref{zscaling} is selected for leptons scattered with $x<0.005$
and $Q^2<1\ \gev^2$, \ie, in the region where strong shadowing is observed in
the DIS cross section. Very similar results were obtained~\cite{e665zdist} in
the non-shadowing region $x>0.03$ and $Q^2>2\ \gev^2$. Apparently, the
shadowing (which naively might be interpreted as a hadronlike behavior of the
virtual photon) does not influence in any way the (absence of) energy loss of
the fast quark in the nucleus. This is what we should expect from the general
space-time picture of DIS -- whether the fast quark is produced in front of or
inside the nucleus it has no time to form a soft gluon cloud before passing
the nucleus.

The trivial $A$-dependence of the hadron distribution in DIS at
high $\nu$ is important in that it establishes a region where the nucleus
behaves in a simple and well understood way. It shows that
even colored particles can penetrate nuclear matter without energy loss, under
proper conditions. Once this is established, one can turn to study the
deviations which appear at low $\nu$. The data~\cite{na2,sal} shows that the
production of hadrons with $z \ge 0.2$ is about 10\% lower on heavy nuclei,
compared to light targets, for $20 \lesim \nu \lesim 80$ GeV. The data at still
lower values of $\nu$ is limited to a data point from an early experiment at
SLAC~\cite{sla} with $\nu \simeq 10$ GeV, which indicates a much stronger
nuclear suppression than the higher $\nu$ data. More data in the
$\nu\lesim 30$ GeV region is needed to map out the effects of
hadronization inside the nucleus.

Our theoretical understanding of energy loss and hadronization effects in a
nuclear medium is still quite limited. The argument used above for a
finite energy loss (hence vanishing fractional energy loss at high $\nu$) is a
direct consequence of the uncertainty principle~\cite{bh}. Detailed
studies of the energy loss in hot QCD matter have been made
in~\cite{gw,bdps,levi}. In Ref.~\cite{bdps} it is concluded that the the energy
loss per unit distance (in an infinitely long medium) is actually proportional
to the square root of the energy of the radiating quark or gluon.

Fits to the observed $A$-dependence of the hadron
distribution~\cite{na2,sal,sla} have been made in a string-inspired
model~\cite{bia,gp}. It was concluded that a good fit required two time scales
in the model. At the `constituent time' the first constituent of the hadron
shows up, and at the `yo-yo time' this constituent finds a partner and forms a
color singlet hadron. In~\cite{kop} it is pointed out that hadrons with a large
fraction of the quark energy, $z \to 1$, are formed early but in a small
configuration. In this regime the effects of color transparency thus become
important.

\section{Color Transparency in DIS} \label{ctdis}

Nuclei are expected to be transparent to fast color singlet system that have
a small transverse size~\cite{ctref}. This phenomenon of {\em color
transparency} (CT) has been the subject of intense experimental and
theoretical interest (see Ref.~\cite{fms} for a recent comprehensive review).
If the transverse size of the (typically \qpair\ or
$qqq$) system is $b$, only gluon interactions with transverse momenta of
\order{1/b} can resolve its color charges, and the nuclear interaction cross
section is expected to be of \order{b^2}. A necessary condition for observing
CT is that the energy $E$ of the compact object is large enough in the nuclear
rest frame, so that its small size $b$ remains frozen during nuclear traversal.
The growth of the transverse size is limited by the transverse velocity
$v_\perp = p_\perp/E \simeq 1/bE$.

\subsection{Compact \qpair\ pairs in the photon} \label{vmct}

DIS provides important tests of CT because the preparation of a transversally
compact \qpair\ object of high energy is particularly simple using photons.
The square of the wave function of a \qpair\ fluctuation in a transversally
polarized photon of virtuality $Q^2$ is~\cite{niza}
\begin{eqnarray}
W_T(\alpha,b)=\frac{6\alpha_{em}}{(2\pi)^2}e_q^2\left\{[1-2\alpha(1-\alpha)]
\epsilon^2 K_1(\epsilon b)^2 \right. \nonumber \\ \left. + m_q^2 K_0(\epsilon
b)^2\right\} \label{wtf}
\end{eqnarray}
where $\alpha$ is the momentum fraction of the quark of charge $ee_q$ and mass
$m_q$, $\alpha_{em}=e^2/4\pi$ and $b$ is the transverse size of the pair.
$K_{0,1}$ are Bessel functions and
\beq
\epsilon^2 = m_q^2+\alpha(1-\alpha)Q^2.  \label{epsq}
\eeq
For small $z$, $K_0(z)\sim -\log(z)$ and $K_1(z) \sim 1/z$, while
for large $z$ the Bessel functions are exponentially small. Thus it can
be seen from \eq{wtf} that the \qpair\ pair has transverse size
$b \lesim 1/\epsilon$. For momentum fractions $\alpha$ of \order{1/2} this
implies, according to the definition (\ref{epsq}), that $b$ is of
\order{1/m_q} or \order{1/Q}, whichever is smaller. This is the
`gluon scattering regime' of \eq{gluon} that we deduced above (in the case
$m_q=0$) from the uncertainty principle.

In the `parton model regime' of \eq{parton} we see that $\epsilon$, and hence
also $1/b$, does not grow with $Q^2$. From \eq{wtf} and the behavior of the
$K_1$ Bessel function we observe that $W_T \propto 1/b^2$ for small
$\epsilon$. The scattering cross-section $\sigma_T$ of the virtual transverse
photon is obtained by multiplying $W_T$ with the \qpair\ interaction
probability $\sigma(b)$,
\beq
\sigma_T= \int_0^1 d\alpha \int d^2 \vec b\, \sigma(b) W_T(\alpha,b).
\label{sigt}
\eeq
With $\sigma(b)\propto b^2$ we find (for $m_q=0$) that the parton model
domain ($\alpha \lesim \morder{1/Q^2}$ or $1-\alpha \lesim \morder{1/Q^2}$)
gives a scaling contribution $\sigma_T \propto 1/Q^2$~\cite{niza,dbp}. All
\qpair\ sizes $b$ contribute in this domain. There is no contribution from
longitudinal photons, whose squared wave function~\cite{niza}
\beq
W_L(\alpha,b)=\frac{6\alpha_{em}}{(2\pi)^2}e_q^2 4Q^2 \alpha^2(1-\alpha)^2
K_0(\epsilon b)^2  \label{wlf}
\eeq
vanishes for $\alpha \to 0,\ 1$.

The gluon scattering regime $\alpha(1-\alpha) \gesim \morder{1/Q^2})$ gives
scaling contributions $\sigma_{L,T} \propto \alpha_s(Q^2)/Q^2$ for both
transverse and longitudinal photons. The $1/Q^2$ suppression in this case is
due to the restriction $b \lesim \morder{1/Q}$, while the factor
$\alpha_s(Q^2)$ results from the hardness of the gluon interaction~\cite{dbp}.

\subsection{$A$-dependence of vector meson production} \label{vmadep}

{}From Eqs. (\ref{wtf},\ref{wlf}) we see that the transverse size of a heavy
quark pair is $b= \morder{1/m_q}$ at low $Q^2$. Early tests of CT
were thus provided by the measurements~\cite{and,aub,sok} of \jpsi\
photoproduction on nuclei, $\gamma A \to J/\psi + X$. Parametrizing the nuclear
target dependence as $\sigma \propto A^\alpha$, E691~\cite{sok} found $\alpha =
0.94 \pm 0.02 \pm 0.02$ in the incoherent region for $p_\perp^2 > 0.15\
\gev^2$. More recently, the NMC collaboration~\cite{nmcpsi} measured $\alpha
= 0.90 \pm 0.03$ for the incoherent elastic process $\gamma A \to J/\psi A$.
These exponents are significantly larger than the $\alpha \simeq 2/3$
expected and observed~\cite{e665} in the photoproduction of light vector
mesons such as the $\rho$.

In the region of the coherent $p_\perp^2$ peak of \jpsi\ production,
E691~\cite{sok} obtained $\alpha = 1.40 \pm 0.06 \pm 0.04$, while
NMC~\cite{nmcpsi} found $\alpha = 1.19 \pm 0.02$. Full transparency would
imply $\sigma_{coh}(A,p_\perp) \propto A^2 \exp(-cA^{2/3}p_\perp^2)$ ($c$
being a constant), resulting in $\alpha = 4/3$ for the $p_\perp$-integrated
cross-section. There is thus rather convincing evidence that the compact
photoproduced $c\bar c$ pairs have a small nuclear reinteraction cross section,
as predicted by CT.

Further evidence for CT has recently come from the measurement by
E665~\cite{e665} of $\gamma^* A\to \rho A$ as a function of the virtuality
$Q^2$ of the photon, as shown in \fref{e665ctfig}. The $A$-dependence
of the incoherent process $(|t'|> 0.1\ \gev^2)$ is seen to be a function of
$Q^2$. For nearly real photons, $\alpha(Q^2=0.212\ \gev^2)= 0.640 \pm 0.030$,
close to 2/3 as expected for a surface dominated soft process. At the highest
measured average virtuality,
$\alpha(Q^2=5.24\ \gev^2)= 0.893 \pm 0.092$, which is consistent with the
value measured for elastic \jpsi\ production~\cite{sok,nmcpsi}. The average
value of the photon energy in this data is about 120 GeV. The nuclear
dependence of
$\rho$ muoproduction has also been measured by the NMC
Collaboration~\cite{nmcrho}. In a sample of events which included both
coherent and incoherent scattering they observed an effective power
$\alpha=1.035 \pm 0.032$, with no significant difference between an average
$Q^2$ of 3.9 $\gev^2$ and 9.6 $\gev^2$. It should be noted that $x\simeq
0.05$ for the high $Q^2$ NMC data, corresponding to a Ioffe length (see
\eq{ioffe})
$L_I = 1/2m_N x \simeq 2$ fm. Hence a significant fraction of the \qpair\
pairs are formed inside the target nucleus, and this fraction is larger for
heavier targets. The measured
$A$-dependence may thus only partly reflect the CT effect. This caveat is
somewhat less serious for the E665 data, for which $x\lesim 0.03$.

\ffig{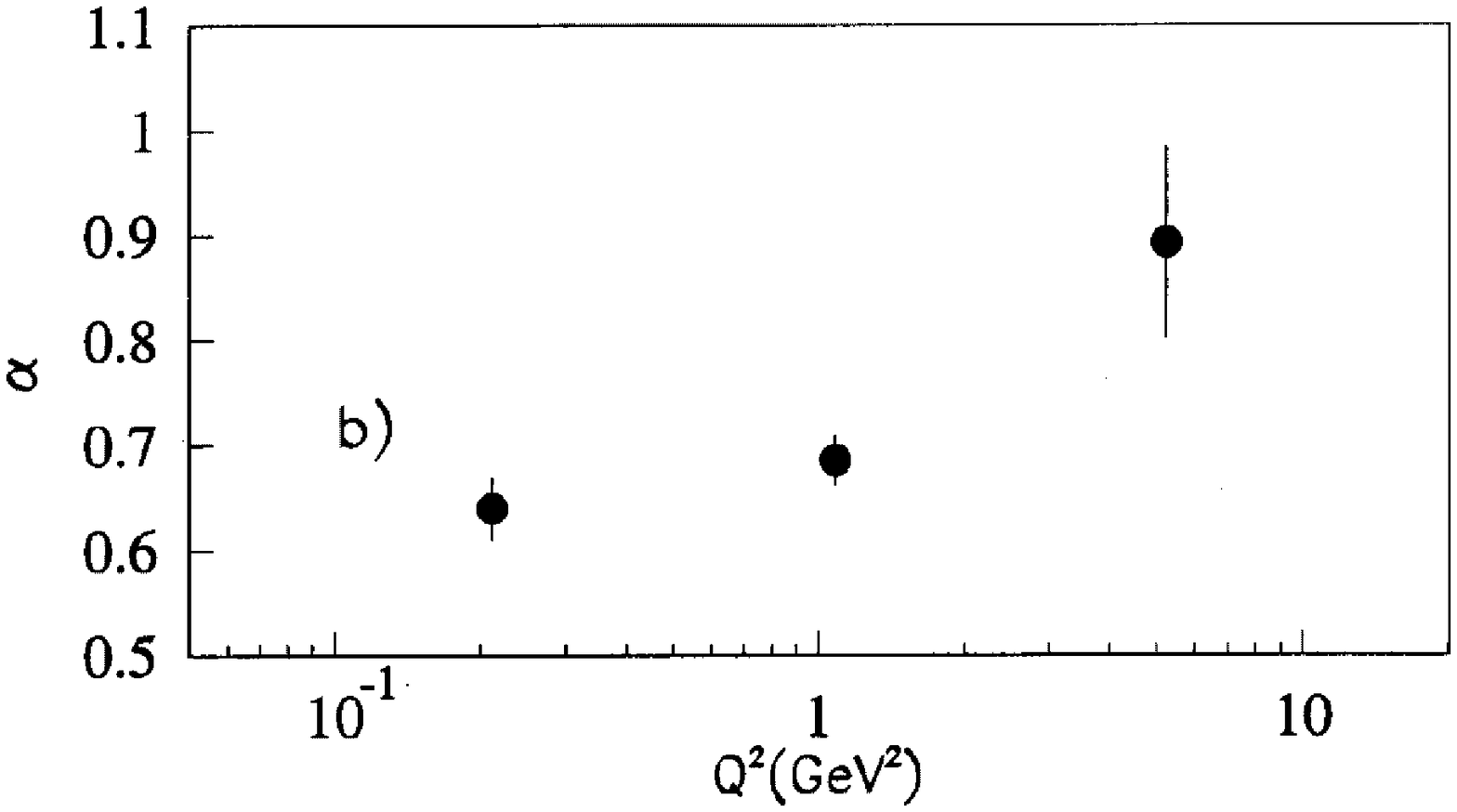}{45mm}{\em The effective exponent $\alpha$ of
the nuclear target dependence in incoherent exclusive $\rho$ muoproduction at
470 GeV~\protect\cite{e665}, $\sigma(\mu A \to \mu \rho A) \propto
A^\alpha$, as a function of the virtuality $Q^2$ of the photon.}{e665ctfig}

\subsection{Comparison with models for vector meson production}

Elastic leptoproduction of vector mesons has been described using an effective
Pomeron exchange model~\cite{dl}, in terms of a constituent quark
picture~\cite{knnz} and using a perturbative two gluon exchange
diagram~\cite{rys,bfghs}. These approaches have many similarities, in
particular they predict (for large $Q^2$) the dominance of longitudinally
polarized vector mesons and a $1/Q^6$ dependence on the virtuality of the
photon.

The data is consistent with some increase with $Q^2$ of the longitudinal to
transverse $\rho$ production ratio, $R=\sigma_L/\sigma_T$. The NMC
Collaboration obtained~\cite{nmcrho} $R=2.0 \pm 0.3$ at $\langle Q^2 \rangle =
6\ \gev^2$, up from $R=-0.38 \pm 0.13_{-0.4}^{+0.9}$ at $Q^2=2\
\gev^2$~\cite{emcrho}. In the HERA energy range, the ZEUS
Collaboration~\cite{zeusrho} measured
$R=1.5^{+2.8}_{-0.6}$ at $\langle Q^2 \rangle = 11\ \gev^2$.

If the $Q^2$ dependence of the $\rho$ production cross section is parametrized
as $1/Q^\beta$, the NMC Collaboration~\cite{nmcrho} obtains $\beta=4.04 \pm
0.14$, with no significant difference between their deuterium, carbon and
calcium targets. The ZEUS~\cite{zeusrho} data give $\beta = 4.2 \pm
0.8^{+1.4}_{-0.5}$. The data suggests that there are important
corrections to the asymptotic $1/Q^6$ behavior expected in the
model calculations, even in the HERA energy range.

The nuclear target $A$-dependence in the Pomeron exchange model would, under
the most simple assumption of a factorizable Pomeron, be independent of $Q^2$
and of the quark mass, since these affect only the virtual photon vertex. As we
have noted above, this is not consistent with the data (nor with CT). Hence, in
a Pomeron exchange picture one is forced to distinguish between the `soft'
Pomeron, familiar from total cross sections and soft hadronic scattering, and
a `hard' Pomeron involved in short distance processes. The `hard'
Pomeron could in fact be nothing but two gluon exchange~\cite{rys,bfghs},
which has CT built into it and so is consistent with the $A$-dependence of the
data, at least at a qualitative level.

The observed dependence of the vector meson cross section on the photon
energy $\nu \simeq s/2m_p$ also demonstrates an important difference between
hard and soft processes. The ZEUS data for $\rho$ production from real
photons~\cite{zeusrhob} shows that the cross section increases only moderately
with $s$, similarly to soft elastic hadron scattering and consistent with a
`soft' Pomeron intercept $\alpha_P(0) \simeq 1.08$. The large $Q^2$ $\rho$
production cross section~\cite{zeusrho} as well as the \jpsi\ cross
section~\cite{zeusjpsi} increases much faster with $s$. In the gluon exchange
model a fast increase is in fact expected. The cross section is (to leading
order in $\log(1/x)$)  predicted~\cite{rys,bfghs} to be proportional to the
square of the gluon structure function, $\sigma \propto [xG(x)]^2$, which
increases rapidly at low $x \simeq Q^2/s$. The predicted $s$-dependence
appears in fact to be somewhat too steep, but is consistent with the data
given the considerable theoretical uncertainties~\cite{zeusrho,zeusjpsi}.
Combined with the observed discrepancy in the $Q^2$-dependence, this may
indicate that only one gluon is effectively hard in the present kinematic
range~\cite{hl}.

It has been suggested~\cite{knnz} that studies of CT for the radial
excitations of vector mesons $(\rho', \psi')$ could reveal interesting
behavior, including an {\em enhancement} of the nuclear production cross
section, as compared to that on free nucleons. The production amplitude is
proportional to the overlap of the \qpair\ wave function in the virtual photon,
\cf\ Eqs. (\ref{wtf}), (\ref{wlf}), and that of the vector meson. The wave
function of the radially excited states has a node, resulting in a cancellation
in the overlap integral. The importance of the cancellation depends on $Q^2$,
which regulates the size of the \qpair\ pair in the photon. It is thus
conceivable that for a suitable value of $Q^2$ the cancellation is almost
complete for $\rho'$ production on free nucleons. Nuclear targets modify the
initial \qpair\ distribution by filtering out large pairs due to
rescattering in the nucleus. This could upset the cancellation and result
in a larger cross section.

More generally, the data on inclusive hadroproduction of \jpsi\ and $\psi'$
illustrates the importance of studying the radially excited states. There are
very considerable discrepancies~\cite{cdfcc} (up to a factor of 50) between the
measured charmonium cross sections and the QCD calculations. Nevertheless, the
ratio of the $\psi'$ to \jpsi\ cross sections is consistent with being
universal~\cite{psirat} for all  beams, targets and reaction kinematics (with
the exception~\cite{na38rat} of nucleus-nucleus collsions). The measured cross
section ratio is
$\sigma(\psi')/\sigma_{dir}(J/\psi) \simeq .24 \pm .05$ for $\pi N$ and $pN$
collisions, and appears to be independent of the target size $A$\footnote{For
production on heavy nuclei the fraction of directly produced \jpsi's (\ie,
those not originating from $\chi_c$ decays) has not been measured, and is
assumed to be the same as that measured on nucleon targets.}. This value is
consistent with expectations based on the \jpsi\ and
$\psi'$ wave functions at the origin. This suggests that the
size of the produced $c\bar c$ pair is small compared to the \jpsi\ and
$\psi'$ wave functions, and that there is little rescattering of the fully
formed charmonium mesons.

The $\psi'$ to \jpsi\ inclusive cross section ratio has also been measured in
muoproduction off the (concrete) absorber in the NMC experiment~\cite{nmcpsi}.
The result, $\sigma(\psi')/\sigma(J/\psi) =0.20 \pm .0.05 \pm 0.07$ can be
directly compared with the ratio quoted above, since the production of $\chi_c$
states from photons should be suppressed. The fact that the ratios are
consistent indicates that the $A$-dependence of the $\psi'$ is similar to that
of the \jpsi\ also in photon induced processes, \ie, the enhancement
scenario~\cite{knnz} apparently does not apply for the charmonium states.

\subsection{Compact $qqq$ configurations in the nucleus} \label{pct}

A complementary way of investigating CT effects is to select compact objects
in the nucleus itself, typically through hard exclusive
scattering~\cite{ctref}. For example, in elastic $ep \to ep$ scattering at
large momentum transfer $Q^2$ it is expected~\cite{blmu} that the only Fock
components of the target proton which contribute are those whose transverse
size is of \order{1/Q}. This follows from general arguments -- the exchanged
photon should scatter coherently over the whole target to avoid a breakup. By
performing the reaction inside a nucleus one hopes to use the nucleus as a
detector to directly measure the size of the scatterer as it recoils through
the nucleus.

It should be noted that at finite energies the contributing Fock states are
not necessarily compact. There is a competing mechanism, the `Feynman'
process~\cite{feyn}, where the initial Fock state has one quark carrying a
large momentum fraction $x \sim 1-1/Q^2$. The electron scatters on this quark
only, while the remaining soft quarks reassociate themselves with the fast one
after the hard scattering so as to form an intact proton moving in the new
direction. Such Fock states can have a large transverse size of \order{1\ {\rm
fm}}, which would upset the CT argument. The importance of this mechanism at
finite $Q^2$ depends on the proton wave function. It is, however, believed to
be subleading at asymptotic energies due to the Sudakov form factor of the
fast quark~\cite{bost}. The form factor expresses the small probability that
a colored object, when given a big momentum transfer, will emit no gluon
radiation, which would imply a breakup of the proton.

An early indication for a CT effect in large $Q^2$ quasielastic $pp$
scattering was obtained from a study of the process $pA \to pp(A-1)$ at
BNL~\cite{bnlct}. Protons of momenta 6, 10 and 12 GeV/c scattered at
$\theta_{cm}^{pp} = 90^\circ$ from $H,\ Li,\ C,\ Al,\ Ca$ and $Pb$, with the
kinematics constrained to correspond to elastic $pp$ scattering and with no
extra particle produced. The nuclear transparency, defined as
\beq
T=\frac{(d\sigma/dt)(pp\ {\rm elastic\ in\ nucleus})}
{(d\sigma/dt)(pp\ {\rm elastic\ in\ hydrogen)}} \label{bnltra}
\eeq
was found to increase with beam momentum until about 10 GeV/c, but then fell
abruptly. In comparison, no energy dependence is expected from a standard
Glauber model calculation, where the struck proton reinteracts with the usual
$pN$ cross section. While the rise of $T$ at lower momenta thus could be a
sign of CT, the interpretation of its decrease above 10 GeV/c is still under
debate. Brodsky and de Teramond~\cite{bdt} have linked this behavior to the
rapid energy dependence seen in the polarization parameter $A_{NN}$ of
$pp$ elastic scattering. They suggest that both phenomena are due to the $c\bar
c$ threshold, which can have a sizable influence on the small
$90^\circ$ cross section. Close to threshold the charm quarks have low momenta
and hence a large transverse size. Ralston and Pire~\cite{rp} observe from the
measured energy dependence of large angle $pp$ elastic scattering that the
protons may at the relevant energies still have a large transverse size
component due to the Landshoff mechanism~\cite{land} of three independent
quark-quark scatterings. For scattering in the nucleus the large proton
components would be filtered away, and thus they argue that the behavior of $T$
is due to the behavior of the denominator rather than of the numerator of
\eq{bnltra}.

A better understanding of the puzzling energy dependence of the
transparency $T$ in $pp$ scattering apparently requires data at higher
energies. If the decrease in $T$ is due to charm threshold, the transparency
should be rapidly restored as the energy is increased, and remain large until
the $b\bar b$ threshold is reached. The Ralston and Pire mechanism, on the
other hand, predicts that the transparency has an oscillating behavior with
energy.

More recently, the NE18 experiment~\cite{ne18} studied CT effects in the $eA
\to ep(A-1)$ process. The energy of the beam electron was $2\ldots 5$ GeV,
$Q^2$ ranged from 1 to 6.8 GeV$^2$ and the targets used were $^2H,\ C,\ Fe$ and
$Au$. The invariant momentum transfer and the recoil proton energy were thus
similar to (although slightly lower than) in the $pp$
experiment~\cite{bnlct}. NE18 defined the transparency as the ratio
of the measured cross section with that estimated using a Plane Wave Impulse
Approximation (PWIA), which neglects any final state interactions of the
struck proton in the nucleus. Effects of the off-shellness and Fermi
momentum of the struck nucleon were included in the PWIA calculation. The
measured transparency was consistent with being independent of $Q^2$ for all
targets (see \fref{ne18transp} for $A=C$). The $A$-dependence of the
transparency was consistent with a Glauber model calculation. The data thus
gave no positive evidence for a CT effect.

\ffig{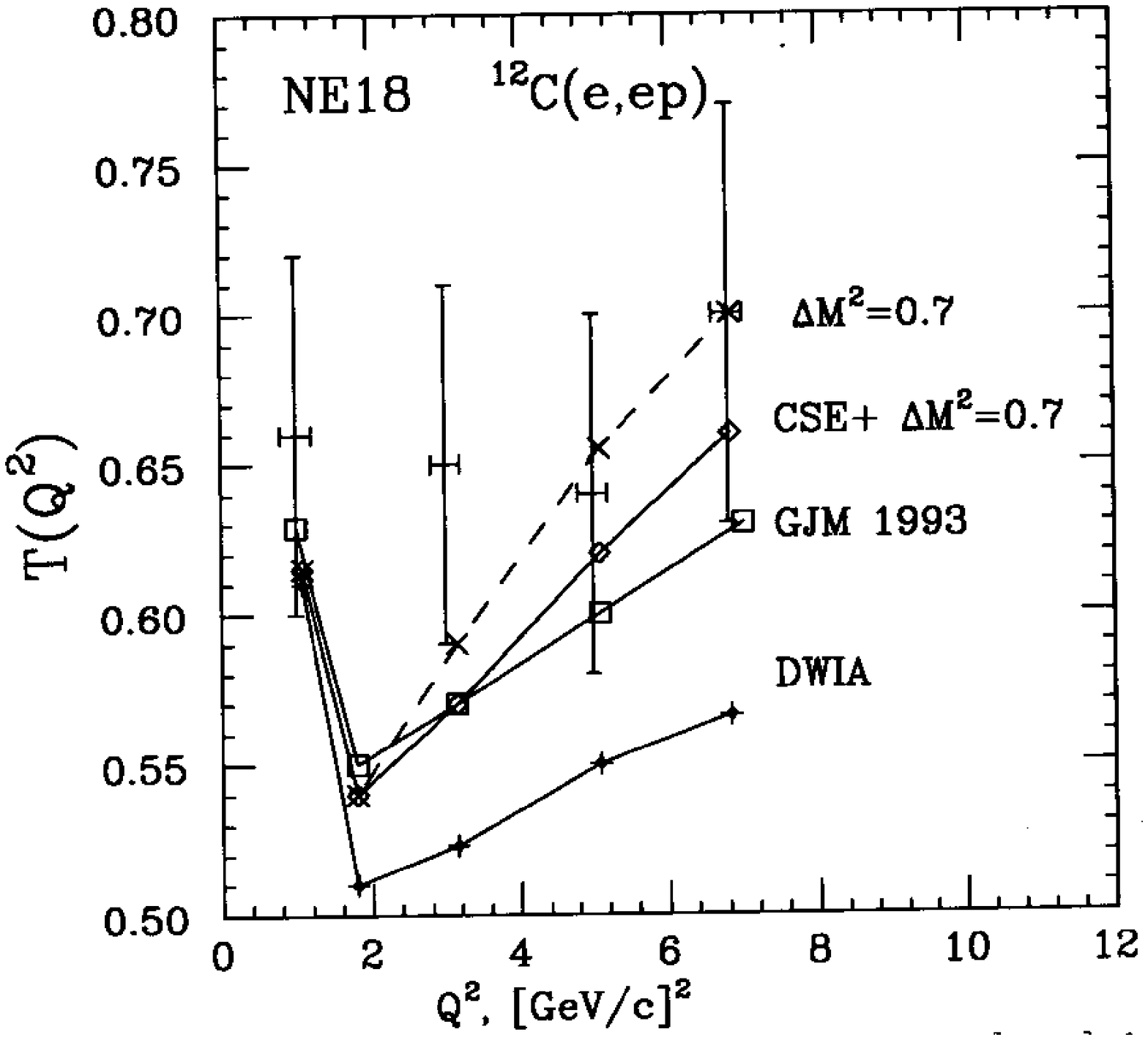}{80mm}{\em Nuclear transparency as measured by
NE18~\protect\cite{ne18} in quasielastic electron scattering on nuclei, $eA\to
ep(A-1)$. The data (error bars) is compared~\protect\cite{fms} with
calculations including the effects of color transparency (CSE, GJM) and with a
Glauber calculation (DWIA). Note the suppressed origin on the vertical
axis.}{ne18transp}

Early theoretical estimates~\cite{cttheor} suggested the possibility of
seeing a modest CT effect. The result of several calculations is
compared~\cite{fms} with the NE18 data in \fref{ne18transp}. As may be seen,
the difference between calculations which include a CT effect (CSE, GJM) and
those without it (DWIA) is marginal. Given the model dependence of the
calculations, no firm conclusions can be drawn. In particular, the same CT
models can fit both the BNL~\cite{bnlct} and NE18~\cite{ne18} data.

\subsection{Comparison of the two types of CT test}

Although the values of $Q^2$ were similar, there are important differences
between the two types of CT test discussed above in sections \ref{vmadep} and
\ref{pct}.
\begin{itemize}
\item {\em Kinematics.}
The vector mesons $(J/\psi, \rho)$ were produced with
energies $\nu$ of \order{100\ \gev}~\cite{sok,nmcpsi,e665} and therefore had
little time to expand before leaving the nucleus. By contrast, the recoil
protons in the BNL~\cite{bnlct} and NE18~\cite{ne18} experiments had energies
of \order{5\ \gev} or less. Hence, their transverse expansion within the
nucleus must be modelled, and any CT effects are diminished.

\item {\em Selection of compact systems.}
In the case of vector meson production, the creation of a compact \qpair\
pair from the virtual photon is rather well understood -- it is governed by
the photon wave function of Eqs. (\ref{wtf}), (\ref{wlf}). On the other hand,
as briefly discussed above, in large angle elastic scattering on nuclei the
size distribution of the contributing proton Fock states for a given $Q^2$ is
uncertain.

\item {\em $A$-dependence of Fock state probabilities.}
The probability distribution of compact \qpair\ pairs in the photon is, as
noted above, governed by the photon wave function and thus independent of the
target size $A$. However, it is not as obvious that the probability of finding
compact $qqq$ states in the nucleus is independent of the nuclear size. In the
CT analysis it must be assumed that this probability is the same as that for
free nucleons. One nucleon shrinks to a small size without the rest
of the nucleons noticing. Nevertheless, properties like the shell structure of
the nucleus must be quite different for the shortlived fluctuations. It is in
principle impossible to determine to which nuclear
energy shell the struck system belonged -- its lifetime $1/Q$ is so short
that the uncertainty in its energy far exceeds the energy spread of
the shell structure (see, however, Ref. \cite{fmss}). The contributions from
nucleons on all shells should be added coherently. Moreover, it is not even
necessary that the $qqq$ system originates from a single nucleon -- compact
states might be formed through the overlap of two or more normal sized
nucleons.
\end{itemize}

\section{High momentum densities in nuclei}  \label{cumsec}

The study of hard exclusive scattering in nuclei is part of an extensive
but still poorly understood area of short distance correlations in nuclei. At
some (low) level of probability, nuclei have dense Fock components, in which
some or all of the quarks and gluons are packed into a small volume. Such
nuclear configurations may have lost much of their nucleon substructure, \ie,
they do not consist of $A$ color neutral $qqq$ subsystems, but can display
normally hidden color degrees of freedom. The dense subsystems give rise to
effects which are kinematically forbidden for scattering on single nucleons.

\subsection{Subthreshold production}
The cross section near threshold for
particle production on nuclei gives an indication of the effective mass and
momentum of the subsystem in the target on which the projectile scatters.
Thus, while the threshold for antiproton production on free nucleons at rest,
$p+p \to \bar p+X$, is $E_{proj}\simeq 6.6$ GeV, the same reaction has been
measured on a copper target, $p+Cu\to \bar p+X$ down to $E_{proj} \lesim 3.0$
GeV~\cite{subt}. If the scattering occured on single nucleons in the copper
nucleus, this would imply Fermi momenta of more than 750 MeV. However, such a
description turns out not to be selfconsistent. Modelling the
nucleon motion by a sufficiently broad Fermi distribution, such that the
$p+Cu$ data could be fitted, resulted~\cite{subt} in an underestimate by a
factor of 1000 for the nucleus-nucleus process $Si+Si \to \bar p +X$ at
$E_{proj}/A \simeq 2.1\ \gev$. Subthreshold production thus appears to involve
scattering on dense subsystems of nucleons or partons, rather than on normal
nucleons.

\subsection{Cumulative production}
It has long been known~\cite{cumrev} that in the scattering of
various projectiles on nuclei, hadrons are produced in the backward direction
with momenta that far exceed the kinematic limit for scattering on single
nucleons at rest. The measured values of the lightcone energy fraction in the
target rest frame for hadrons produced near 180$^\circ$, $x=(E-p_L)/m_A$,
reach up to values of $x\simeq 4$ for protons in $pA$ collisions~\cite{boya},
suggesting that at least four nucleons in the nucleus have been involved in the
scattering. These cumulative phenomena are only weakly dependent on the
projectile type and energy, having been observed for hadron, photon, neutrino
and nuclear projectiles of momenta from 1 to 400 GeV. This strongly suggests
that the cumulative phenomena are related to the nuclear target wave function.
In $A_pA_t$ nucleus-nucleus collisions the dependence of the cumulative cross
section on the atomic number of the projectile $A_p$ and target $A_t$ has been
found~\cite{geag} to scale like $A_p^{2/3}A_t^{4/3}$, consistent with a soft
(surface dominated) projectile scattering but a faster than volume increase
with the target size. There is evidence~\cite{boya} that the backward produced
hadrons have large transverse momenta, suggesting that they originate from a
target subsystem of small transverse size.

\subsection{DIS at $x \gesim 1$}.
A direct way of observing phenomena that only can occur in nuclei
is DIS at $x>1$. Present data~\cite{day} at the largest $x \simeq 2$ is
limited to low values of the photon energy $\nu = \morder{300 \dots 500\
{\rm MeV}}$, and is thus not really in the `deep inelastic' region.
Nevertheless, various scaling phenomena have been studied~\cite{fsds}. For $x
\simeq 1$ and $Q^2 \lesim 3\ \gev^2$ the reaction is quasielastic, and scaling
is observed in a variable $y$ related to the Fermi motion of the struck
nucleon. The $y$-scaling is found~\cite{arri} to break down at higher values
of $Q^2$, where the scattering becomes inelastic and presumably occurs off
quarks rather than nucleons. As shown in \fref{arrfig3} a rough scaling
in the whole $Q^2$ range is, however, observed in terms of the Nachtmann
scaling variable $\xi=2x/[1+(1+4m_p^2x^2/Q^2)^{1/2}]$ (which takes into
account target mass effects), suggesting a duality between the quasielastic
and inelastic processes.

\ffig{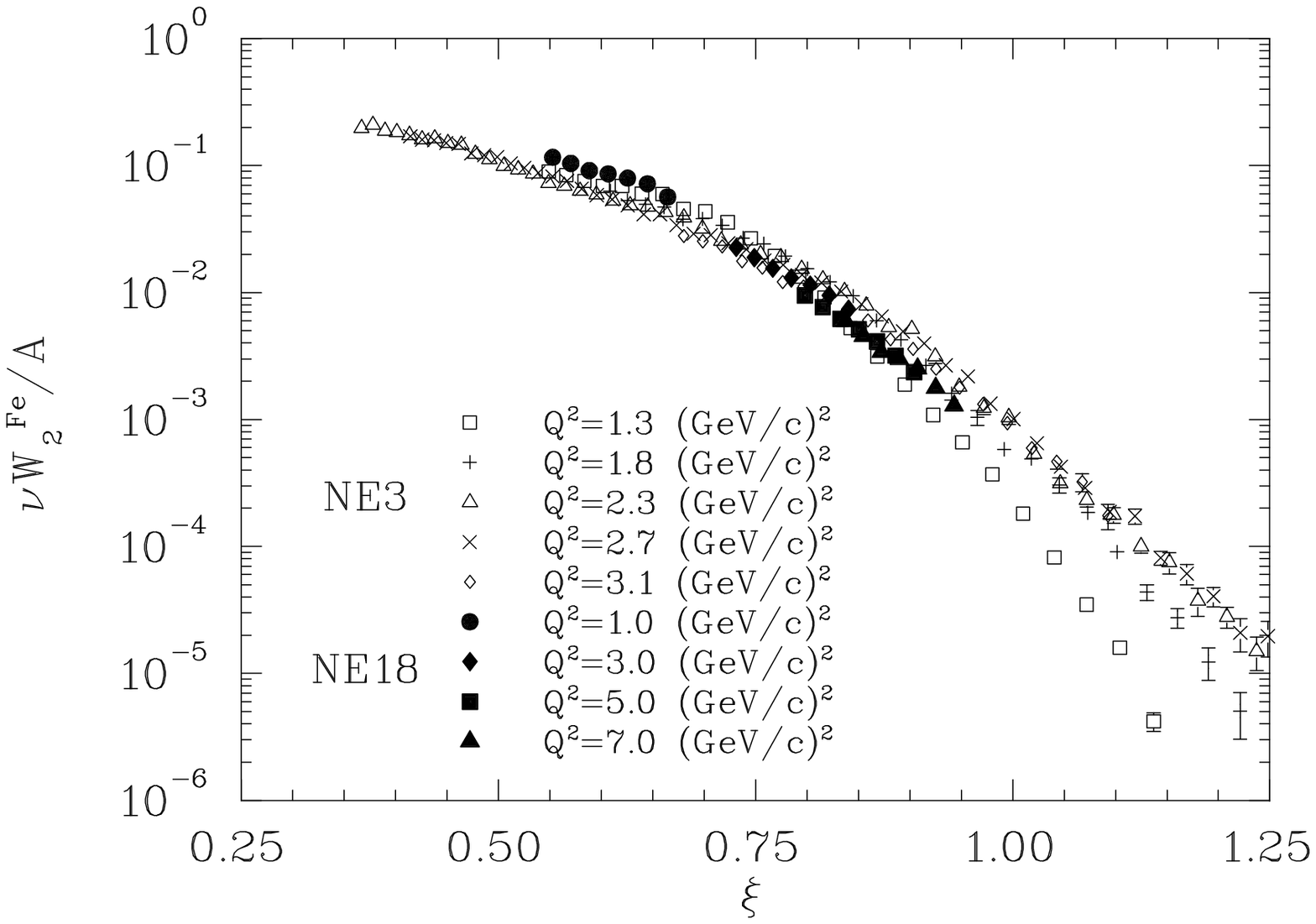}{60mm}{\em $\nu W_2^{Fe}$ as measured at large $x$ by the NE3
and NE18 experiments~\protect\cite{arri}, shown as a function of the Nachtmann
scaling variable $\xi$ for a range of $Q^2$ values.}{arrfig3}

The behavior of hard QCD processes at large values of $x$, where higher twist
processes dominate and several partons scatter coherently, have
been studied qualitatively in Ref.~\cite{bhmt}. The general features agree with
what is observed, \eg, in the cumulative process of backward hadron production.
In particular, the scattering of the projectile becomes softer, since it occurs
off the soft stopped partons that contributed their momenta to the large $x$
subsystem.

\section{Summary}

This presentation has perforce covered only a minor part of the rich
phenomena that occur in scattering on nuclei, and the selection has been
heavily influenced by my own limited knowledge of the field. Nevertheless, I
am convinced that light can be thrown on many fundamental
questions through DIS on nuclei. Here `DIS' is taken to mean not only the
fully inclusive process $eA \to eX$, but more generally hard scattering on
nuclei initiated by real or virtual photons. The topics can be roughly divided
into two classes.
\begin{itemize}
\item {\em Space-time development of hard scattering.} Secondary (soft)
scattering in the extended nucleus gives information about the development of
the partonic subsystems before and after the hard scattering. The nucleus
acts as a `femtovertex' detector. The {\em shadowing} of DIS at small $x$
indicates the transverse size of the system formed by the virtual photon. The
(absence of) {\em energy loss} of the fast quark produced by the virtual
photon shows the slow development of hadronization. The {\em color
transparency} observed in the production of \jpsi\ and $\rho$ mesons indicates
that the quark pairs created by highly virtual photons have a small size
compared to the confinement scale.

\item {\em Short-range correlations in nuclei.} Perturbative QCD should
correctly describe phenomena taking place over distances $\ll$ 1 fm. Nuclei
provide an extensive testing ground for the theory, and
can give us valuable insights into the applicability of QCD. The
experimental signals of {\em subthreshold production, cumulative processes}
and {\em DIS at $x>1$} are still poorly understood theoretically. High
sensitivity experiments are needed to properly study the rare Fock states of
nuclei.
\end{itemize}

Real and virtual photons are the most precise tools we have for exploring the
fine structure of matter. It is very important that they can be brought to
bear also on nuclear targets. With facilities such as CEBAF, HERMES and
ELFE the future prospects look promising.

 \begin{center} {\large\bf Acknowledgements} \end{center} It is a pleasure
to thank the organizing committee for inviting me to present
this review. I have benefitted from discussions on hard scattering in
nuclei with many colleagues, and in particular with Stan Brodsky, Boris
Kopeliovich, Mark Strikman, Wai-Keung Tang, Ramona Vogt and Mikko V\"anttinen.

\end{document}